\documentclass[fleqn,usenatbib]{mnras}
\usepackage{newtxtext,newtxmath}

\usepackage[T1]{fontenc}
\usepackage{ae,aecompl}

\usepackage{graphicx}	% Including figure files
\usepackage{amsmath}	% Advanced maths commands
\usepackage{amssymb}	% Extra maths symbols
\title[IR-radio correlation of spheroid and disc galaxies]{The infrared-radio correlation of spheroid- and disc-dominated star-forming galaxies to z $\sim$ 1.5 in the COSMOS field}

\author[D. Cs. Moln\'{a}r et al.]{
D\'{a}niel Cs. Moln\'{a}r,$^{1, 2}$\thanks{E-mail: d.molnar@sussex.ac.uk}
Mark T. Sargent,$^{1}$
Jacinta Delhaize,$^{2}$
Ivan Delvecchio,$^{2}$
\newauthor
Vernesa Smol\v{c}i\'{c},$^{2}$
Mladen Novak,$^{2}$
Eva Schinnerer,$^{3}$
Giovanni Zamorani,$^{4}$
Marco Bondi,$^{5}$
\newauthor
Noelia Herrera-Ruiz,$^{6}$
Eric J. Murphy,$^{7}$
Eleni Vardoulaki,$^{8}$
Alexander Karim,$^{8}$
\newauthor
Sarah Leslie,$^{3}$
Benjamin Magnelli,$^{8}$
C. Marcella Carollo,$^{9}$
Enno Middelberg$^{6}$
\\
$^{1}$Astronomy Centre, Department of Physics \& Astronomy, University of Sussex, Brighton, BN1 9QH, England\\
$^{2}$Department of Physics, Faculty of Science, University of Zagreb,  Bijeni\v{c}ka cesta 32, 10000  Zagreb, Croatia\\
$^{3}$MPI for Astronomy, K\"{o}nigstuhl 17, D-69117 Heidelberg, Germany\\
$^{4}$INAF--Osservatorio Astronomico di Bologna, via P. Gobetti 93/3, 40129 Bologna, Italy\\
$^{5}$INAF--Istituto di Radioastronomia, via P. Gobetti 101, 40129 Bologna, Italy\\
$^{6}$ Astronomisches Institut, Ruhr-Universit\"{a}t Bochum, Universit\"{a}tstrasse 150, 44801 Bochum, Germany\\
$^{7}$National Radio Astronomy Observatory, 520 Edgemont Road, Charlottesville, VA 22903, USA\\
$^{8}$Argelander Institute for Astronomy, University of Bonn, Auf dem H\"{u}gel 71, D-53121 Bonn, Germany\\
$^{9}$Department of Physics, Institute for Astronomy, ETH Z\"{u}rich, CH-8093 Z\"{u}rich, Switzerland
}

\date{Accepted XXX. Received YYY; in original form ZZZ}

\pubyear{2017}

\begin{document}
\label{firstpage}
\pagerange{\pageref{firstpage}--\pageref{lastpage}}
\maketitle

\begin{abstract}
Using infrared data from the Herschel Space Observatory and Karl G. Jansky Very Large Array (VLA) 3 GHz observations in the COSMOS field, we investigate the redshift evolution of the infrared-radio correlation (IRRC) for star-forming galaxies (SFGs) we classify as either spheroid- or disc-dominated based on their morphology. The sample predominantly consists of disc galaxies with stellar mass ${\gtrsim}10^{10}\,M_{\odot}$, and residing on the star-forming main sequence (MS). After the removal of AGN using standard approaches, we observe a significant difference between the redshift-evolution of the median IR/radio ratio $\overline{q}_{\mathrm{TIR}}$ of (i) a sample of ellipticals, plus discs with a substantial bulge component (`spheroid-dominated' SFGs) and, (ii) virtually pure discs and irregular systems (`disc-dominated' SFGs). The spheroid-dominated population follows a declining $\overline{q}_{\mathrm{TIR}}$ vs. $z$ trend similar to that measured in recent evolutionary studies of the IRRC. However, for disc-dominated galaxies, where radio and IR emission should be linked to star formation in the most straightforward way, we measure very little change in $\overline{q}_{\mathrm{TIR}}$. This suggests that low-redshift calibrations of radio emission as an SFR-tracer may remain valid out to at least $z\,{\simeq}\,1\,{-}\,1.5$ for pure star-forming systems. We find that the different redshift-evolution of $q_{\rm TIR}$ for the spheroid- and disc-dominated sample is mainly due to an increasing radio excess for spheroid-dominated galaxies at $z\,{\gtrsim}\,$0.8, hinting at some residual AGN activity in these systems. This finding demonstrates that in the absence of AGN the IRRC is independent of redshift, and that radio observations can therefore be used to estimate SFRs at all redshifts for genuinely star-forming galaxies.
\end{abstract}
\begin{keywords}
galaxies: evolution -- radio continuum: galaxies -- infrared: galaxies
\end{keywords}

%%%%%%%%%%%%%%%%%%%%%%%%%%%%%%%%%%%%%%%%%%%%%%%%%%

%%%%%%%%%%%%%%%%% BODY OF PAPER %%%%%%%%%%%%%%%%%%

\section{Introduction}
\label{sect::intro}

Observations show that the total infrared and 1.4 GHz radio continuum luminosities of local galaxies are tightly correlated \citep{kruit71, kruit73, Jong85, Helou85, condon92, yun01}. This so-called infrared-radio correlation (IRRC) was found to be linear over at least three order of magnitudes in luminosity since the peak epoch of star formation \citep[e.g.][]{sajina08, murphy09}. Thus it has been used to e.g. identify radio-loud AGN \citep[e.g.][]{donley05, norris06, park08, delmoro13}, and to estimate the distances and temperatures of high-redshift submillimetre galaxies \citep[e.g.][]{carilli99, chapman05}. It also enables the calibration of radio luminosities as dust-unbiased, high angular resolution star formation rate (SFR) tracers \citep[e.g.][]{condon92, bell03, murphy11, murphy12, delhaize17, davies17}. Future deep radio continuum surveys aiming to obtain a full census of dust-obscured star formation across cosmic time, even in crowded environments such as groups and clusters, will thus heavily rely on the measured IRRC to achieve their science goals. With the advent of new, high-sensitivity radio instruments (such as LOFAR, MeerKAT, ASKAP and the Square Kilometre Array) it is timely to study the redshift evolution and higher order dependencies of the IRRC, which contribute to, e.g., the scatter of the relation, in more detail.

Observationally it has been challenging to refine existing work on the IRRC due to the lack of sufficiently sensitive radio and infrared data. \citet{sargent10} showed that solely radio or infrared selected flux-limited surveys introduce a bias that artificially produces an evolution in the IRRC. To overcome this, they used flux limits for non-detections at either of these wavelength, and constrained the median infrared radio ratio with double censored survival analysis. They found no significant evolution in the IRRC out to $z$\,$\sim$\,1.5 using VLA imaging of the Cosmological Evolution Survey \citep[COSMOS;][]{scoville07} field at 1.4 GHz \citep{schinnerer07, schinnerer10}. Many other authors \citep[e.g.][]{garrett02,appleton04, ibar08, garn09, jarvis10,mao11,smith14} have also found no significant evidence for evolution of the IRRC up to $z$\,$\sim$\,3.5. However, a series of recent studies relying on the full far-IR coverage provided by the Herschel Space Observatory \citep{pilbratt10}, have found evidence for a changing IRRC across cosmic times \citep[but see also][]{pannella15}. \citet{magnelli15} performed a stacking analysis of Herschel, VLA and Giant Metre-wave Radio Telescope (GMRT) radio continuum data to study the variation of the IRRC out to z $<$ 2.3. They found a moderate, but statistically-significant evolution with $(1+z)^{-(0.12\pm0.04)}$. Matching the depth of Herschel data with the sensitivity of LOFAR \citep{harleem13}, \citet{calistro-rivera17} measured a consistent redshift dependency of $(1+z)^{-(0.15\pm0.03)}$ for star forming galaxies out to z $\sim$ 2.5. Most recently, \citet[][D17 henceforth]{delhaize17} presented new evidence for a similar declining trend in the infrared/radio ratio out to z $<$ 5 using new, deep 3 GHz VLA images and Herschel FIR fluxes in the COSMOS field.

From a theoretical perspective, star-formation (SF) is thought to be the link between infrared and radio emission. Young, massive (>8\,$M_{\odot}$) stars produce UV photons, that are mostly absorbed and re-emitted by the surrounding dust at far-infrared (FIR) wavelengths. Radio emission at low rest-frame GHz frequencies predominantly represents non-thermal synchrotron radiation emitted by relativistic cosmic ray (CR) electrons that move through galactic magnetic fields as they are accelerated by supernovae remnants \citep[e.g.][]{condon92}. The IRRC thus arises if CR electrons radiate away all their energy before escaping the galaxy and if the interstellar medium (ISM) is optically thick in UV, reprocessing all the UV starlight into FIR emission. This so-called calorimetry theory was first proposed by \citet{voelk89}. Since then several studies have pointed out its shortcomings and provided alternative, more complex explanations \citep[e.g.][]{helou93,bell03,lacki10a,schleicher2013}, however the exact physical processes driving the relation remain unclear. Predictions for the redshift evolution of the IRRC are also often conflicting. On the one hand, an increase of infrared/radio flux ratios is expected \citep[e.g.][and references therein]{murphy09}; rather than undergoing synchrotron cooling, at higher redshift CR electrons should lose more and more energy through inverse Compton (IC) scattering off photons of the cosmic microwave background (CMB) as the CMB energy density increases with redshift, thereby violating electron calorimetry. On the other hand, \citet{lacki10b} argues that this effect can be compensated by other IC loss effects (e.g. ionization, bremsstrahlung) preserving radio luminosity, especially in starbursts, hence keeping the relation constant up to at least $z\,{\sim}\,1.5$.

An approach to observationally determine which physical processes regulate the IRRC is to identify what factors contribute to its $\sim$ 0.26 dex scatter \citep{bell03}. One would expect that, e.g., a non-SF related warm cirrus component at IR wavelengths or excess radio emission due to low level AGN activity, both phenomena more common in ``red and dead'' early-type galaxies, would cause deviations from the median IR/radio ratio. Finding such differences motivated this work, which is a direct follow-up of D17. Utilising the wealth of ancillary data available in the COSMOS field, here we will study the IRRC's evolution in disc- and spheroid-dominated SFGs.

Throughout this paper, we use a flat $\Lambda$CDM cosmology with $\Omega_M = 0.3$ and $\mathrm{H_0 = 70 \, km\, Mpc^{-1}\, s^{-1}}$. Star formation rates and stellar mass values reported assume a Chabrier initial mass function (IMF).

\section{Data}
\label{sect:data}

In order to investigate the dependence of the IRRC on galaxy morphology, we used a combination of Herschel and VLA 1.4 and 3 GHz data with added structural information from the Zurich Structure \& Morphology Catalog\footnote{Full catalogue with description is available at \url{http://irsa.ipac.caltech.edu/data/COSMOS/tables/morphology/cosmos_morph_zurich_colDescriptions.html}.}. \citet{sargent10} showed that considering only IR- or radio-selected samples of star-forming galaxies (SFGs) tends to, respectively, over- or underestimate average IR/radio ratios, with the typical offset between such samples being $\sim$0.3\,dex. This is approximately the same as the intrinsic scatter of the IRRC itself \citep[e.g. ][]{yun01}, hence an accurate analysis of the relation should account for flux limits in both wavelength regimes. To overcome this bias we use a jointly-selected (i.e. both radio and IR) sample of galaxies. The following section briefly outlines the sample construction process. More details on all aspects of the summary provided in Sect. \ref{sect::joint-samp} can be found in D17.

\subsection{Jointly-selected parent catalogue}
\label{sect::joint-samp}

We use the jointly-selected sample of D17 as our parent catalogue. It is the union of radio- and IR-selected samples in the 2 $\mathrm{deg^2}$ COSMOS field.

The IR-selected galaxy sample was constructed from a prior-based catalogue of Herschel flux measurements in the COSMOS field. 100 and 160 $\mu$m Herschel Photodetector Array Camera \citep[PACS; ][]{poglitsch10} data are from the PACS Evolutionary Probe \citep[PEP; ][]{lutz12}. 250, 350 and 500 $\mu$m maps are provided by the Herschel Multi-tiered Extragalactic Survey \citep[HerMES;][]{oliver10}. Their prior positions from the 24\,$\mu$m Spitzer \citep{lefloch09} MIPS catalogue were matched to the COSMOS2015 photometric catalogue \citep{laigle16} containing both photometric and spectroscopic redshift information. D17 selected sources with $\geq$5\,$\sigma$ detections in at least one Herschel band, in order to obtain an IR-selected catalogue that is comparable to the radio-selected sample (see below) in terms of SFR-sensitivity. This IR-selected catalogue contains 8,458 sources. 

Our radio-selected catalogue is based on the imaging from the 3 GHz VLA-COSMOS Large Project. The VLA-COSMOS 3\,GHz catalogue \citep{smolcic17a} contains $\sim$11,000 sources down to $S/N$\,=\,5 (the typical sensitivity is 2.3 $\mu$Jy beam$\mathrm{^{-1}}$ over most of the 2 $\mathrm{deg^2}$ area). In order to obtain rest-frame spectral energy distributions (SEDs) and calculate luminosities, reliable redshifts are required. \citet{smolcic17b} thus assigned optical/near-IR (NIR) counterparts in unmasked regions (i.e. avoiding bright, saturated stars) of the COSMOS field where high-quality photometric data were available from the COSMOS2015 photometry catalogue \citep[presented in ][]{laigle16}. The sample of radio-selected sources with optical counterparts and known photometric or spectroscopic redshifts initially contains 7,729 objects. However, due to the 3 GHz mosaic's high angular resolution, some low surface brightness sources were below the 5\,$\sigma$ detection limit of this radio catalogue. To correct for this resolution bias, D17 searched for additional detections in 3 GHz maps convolved to lower resolutions of up to 3 arcsec and at the positions of IR-selected sources with no 3 GHz counterparts in the 0.75 arcsec mosaic. This yielded an additional 428 sources. We also added 1.4 GHz radio fluxes from the VLA-COSMOS catalogue \citep{schinnerer07} to 27 IR-selected sources with no 3 GHz counterpart. Thus the final radio-selected sample consists of 8,184 sources.

The resulting jointly-selected sample contains 4,309 objects that were detected at both radio and IR wavelengths, 3,875 sources that have only radio detections and 4,149 sources with only IR fluxes measured. In the latter two cases we used 5\,$\sigma$ upper flux limits for the non-detections to help constrain the median infrared radio ratio. 37 \% of the galaxies have spectroscopic redshifts in both the radio- and the IR-selected samples.

\subsection{Morphologically-selected sub-samples}
\label{sect::morph_cat}

To add morphological information to the jointly-selected sample of D17, we cross-matched it with the Zurich Structure \& Morphology Catalog. This catalog provides a classification of galaxies into different  categories based on the ZEST (Zurich Estimator of Structural Types) algorithm \citep{scarlata07}. ZEST uses five nonparametric structural diagnostics (asymmetry, concentration index, Gini coefficient, second-order moment of the brightest 20 \% of galaxy pixels, and ellipticity) measured on Hubble Space Telescope (HST) Advanced Camera for Surveys (ACS) I-band (F814W) images \citep{koekemoer07} to morphologically classify sources. After carrying out a principal component (PC) analysis to reduce the number of parameters while retaining most of their information content, ZEST  uses a 3D classification grid to define three main galaxy types: elliptical (type 1), disc (type 2) and irregular (type 3) objects. Type 2 was then further divided into four bins (i.e. 2.3, 2.2, 2.1 and 2.0), guided by the S\'{e}rsic index $n$ of galaxies in the `disc' class \citep{sargent07}. These sub-classes reflect an increasing prominence of the bulge component from type 2.3 to 2.0, i.e. type 2.3 are pure disc galaxies, type 2.0 are strongly bulge-dominated discs, while types 2.1 and 2.2 have intermediate bulge-to-disc ratios. The application of ZEST to a sample of low-redshift galaxies from \citet{frei96} with RC3 classifications \citep{deVac91}, showed that type 1 objects are mostly classified as Hubble type E, type 2.0 corresponds to S0 - Sab galaxies, type 2.1 mainly consists of Sb - Scd systems, type 2.2 sources are split between Sb-Scd and Sd and later types, and type 2.3 maps into Sd discs or even later RC3 types \citep[for detailed distributions see Fig. 6 in][]{scarlata07}.

\begin{table*}
	\centering
	\caption{Morphological distribution of galaxies before (top) and after (bottom) the exclusion of AGN (see Sect. \ref{sect::agn_id}) from the radio- and infrared-selected samples. Morphological categories are: ellipticals (ZEST type 1); bulge-dominated discs discs (ZEST type 2.0/2.1); disc-dominated discs (ZEST type 2.2/2.3); irregulars (ZEST type 3).}
   	\label{tab:distr}
	\begin{tabular}{cccccc} % four columns, alignment for each
		\hline
		 \multicolumn{6}{c}{\it Full sample} \\
		\hline
		 & \multicolumn{2}{c}{spheroid-dominated}  & \multicolumn{2}{c}{disc-dominated} & Total\\
		 & ellipticals & bulge-dominated discs & disc-dominated discs & irregulars & \\
		\hline
		radio-selected & 503 & 1461 & 1420 & 595 & 3979\\
		IR-selected & 157 & 1305 & 2098 & 802 & 4362\\
        jointly-selected & 545 & 2010 & 2544 & 941 & 6040\\
		\hline \hline
		 \multicolumn{6}{c}{\it Star-forming sample} \\
		\hline
		 & \multicolumn{2}{c}{spheroid-dominated}  & \multicolumn{2}{c}{disc-dominated} & Total\\
		 & ellipticals & bulge-dominated discs & disc-dominated discs & irregulars & \\
		\hline
		radio-selected & 130 & 905 & 1248 & 514 & 2797\\
		IR-selected & 125 & 1142 & 1971 & 727 & 3965\\
        jointly-selected & 168 & 1409 & 2323 & 836 & 4736\\
		\hline
	\end{tabular}
\end{table*}

\begin{figure*}
\includegraphics[width=0.86\textwidth]{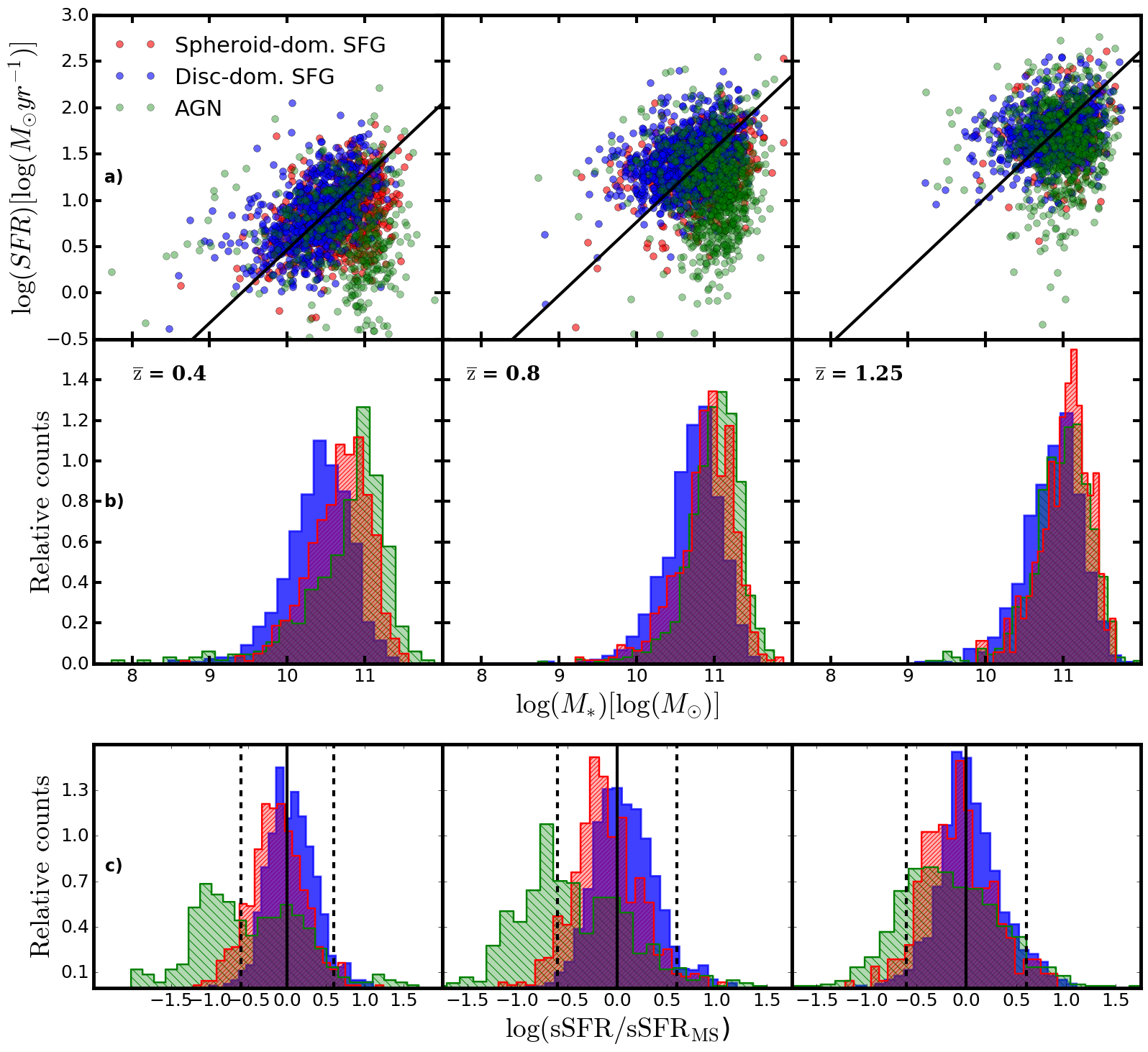}
\caption{\textbf{a)} Spheroid- and disc-dominated SFGs and AGN in the stellar mass -- SFR plane in three redshift slices. The mean redshift $\overline{z}$ is given for each redshift bin. Black line represent the average locus of the star-forming main sequence (parametrised as in the appendix of \protect\citealp{sargent14}) at the $\overline{z}$ of each panel. \textbf{b)} Normalised (equal area) stellar mass distributions of spheroid- and disc-dominated SFGs and AGN. Spheroid-dominated SFGs are systematically more massive across the whole studied redshift range. \textbf{c)} Normalised distribution of sSFR-offsets from the star-forming main sequence for spheroid- and disc-dominated SFGs. Vertical dashed lines show 0.6 dex offsets above and below the SF MS. The mean sSFR of disc-dominated (spheroid-dominated) SFGs tends to lie above (below) the average MS locus at all redshifts. AGN mainly occupy the quiescent regime below $z \sim 1$.}
\label{fig::ssfr_rat}
\end{figure*}

The Zurich Structure \& Morphology Catalog was position-matched to the optical positions of the jointly-selected sample with a search radius of 0.6 arcsec. 7,973 (65\%) sources in the jointly-selected sample (see Sect. \ref{sect::joint-samp}) were covered by HST/ACS and satisfy the $i_{\rm AB}$\,=\,24\,mag selection limit of the Zurich Structure \& Morphology Catalog. 6,723 of these had a counterpart with morphological classification in the morphology catalogue, resulting in an 84 \% matching rate. At redshifts $z$\,>\,1.5 the F814W filter starts to sample rest-frame UV ($<$325\,nm) emission, and image signal-to-noise and galaxy angular sizes in general become too small to ensure a robust morphological classification. We thus apply an upper redshift cut of $z$\,=\,1.5 to our sample. Due to the small volume sampled locally by the $\sim$1.6\,deg$^2$ field observed with HST/ACS, we excluded sources with $z$\,<\,0.2. Our final sample contains 6,072 galaxies with morphological distribution as summarized in Table \ref{tab:distr}. Our `spheroid-dominated' sample includes ZEST types 1 (i.e. predominantly elliptical galaxies) and types 2.0/2.1 (i.e. disc galaxies with a prominent bulge component). Our `disc-dominated' galaxy sample includes ZEST types 2.2/2.3 (i.e. disc-dominated) spiral/disc galaxies and type 3 (irregular galaxies). We note that both morphological groups contain mainly disc galaxies (see Table \ref{tab:distr}). After the exclusion of AGN (see Sect. \ref{sect::agn_id}), the remaining star-forming sample predominantly lies on the main sequence (MS) of star-forming galaxies (see Fig. \ref{fig::ssfr_rat}a), consistently with previous studies \citep[e.g.][]{wuyts11}, as illustrated by the specific star formation rate (sSFR) distributions in Fig. \ref{fig::ssfr_rat}c. The stellar masses used here are from multi-band SED modelling with \textsc{magphys}, and SFR values were derived from the fitted IR luminosities (see Sect. \ref{sect::ir_lum} for details). We will henceforth qualitatively refer to these morphological categories as `{\it disc-dominated star-forming galaxies}' and `{\it spheroid-dominated star-forming galaxies}', respectively. We also note that disc- and spheroid-dominated SFGs have a tendency to lie slightly above and below the main sequence locus, with mean offsets from the MS of 0.06 and -0.10, respectively. Furthermore, spheroid-dominated galaxies on average have higher stellar masses at all redshifts considered here (see Fig. \ref{fig::ssfr_rat}b). 
Assuming that the radio and far-IR data used to select our sample are due to star formation alone, our sample is SFR-selected, and in deriving all results reported in the following we also include galaxies that lie below the mass-completeness threshold at 10$^{10.4}\,M_{\odot}$ \citep{laigle16}. We have ascertained that all our results remain unchanged within 1\,$\sigma$ if we restrict the analysis to the mass-complete regime.\\
Fig. \ref{fig::frac_vs_z}a shows the relative abundance of spheroid- to disc-dominated SFGs as a function of redshift. Over the redshift range 0.2\,<\,$z$\,<\,1.5 the overall disc galaxy population transitions from consisting mainly of by disc-dominated objects at high-$z$ to a more even split between morphological types at low redshift (see Fig. \ref{fig::frac_vs_z}a). Our sample is hence representative of the COSMOS disc galaxy population as a whole for which this trend was already discussed in, e.g., \citet{scarlata07} and \citet{oesch10}.

\subsection{AGN identification}
\label{sect::agn_id}

Our aim is to test the relation between radio synchrotron emission and SFR for different galaxy populations. Identifying and removing potential AGN host galaxies from our sample is hence crucial as both their IR and radio fluxes could include AGN-related contributions. Sources in the jointly-selected catalogue were flagged in D17 as likely AGN hosts if at least one of the following criteria was met:
\begin{enumerate}
\item the source shows a power-law like emission in the mid-IR, i.e. its IRAC colours satisfy the criteria of \citet[][their eqs. (1) \& (2)]{donley12},
\item an X-ray detection in the combined maps of the Chandra-COSMOS and COSMOS Legacy surveys \citep{elvis09,civano12,civano16,marchesi16} with [0.5-8] keV X-ray luminosity $L_X\,>\,10^{42}$ erg\,s$^{-1}$ \citep[as in][]{smolcic17b},
\item SED fitting reveals the presence of a statistically significant AGN component based on a $\chi^2$ comparison between fits without and with an AGN contribution of freely variable amplitude \citep{delvecchio14},\footnote{A multi-component SED fit was performed using \uppercase{sed3fit} \citep{berta13}, publicly available at \url{http://cosmos.astro.caltech.edu/page/other-tools}.}
\item the source is not detected in any Herschel bands with a signal-to-noise ratio of at least 5, and has red optical rest-frame colours ($M_{NUV}\,-\,M_r$)\,>\,3.5 \citep{smolcic17b}.
\end{enumerate}
For further details see D17 and \citet{delvecchio17}. 1,304 galaxies in our sample are flagged as AGN ($\sim$38\% of spheroid-dominated and $\sim$9\% of disc-dominated galaxies) based on these criteria. The final sample of jointly IR- and radio-selected star-forming galaxies with a morphological classification from the Zurich Structure \& Morphology Catalog consists of 4,736 sources. Table \ref{tab:distr} summarizes how these star-forming galaxies split into distinct morphological classes. We will consider only the star-forming population for the rest of our analysis, unless stated otherwise.

\section{Results}

\subsection{Derivation of radio luminosities}

Radio luminosities were derived using
\begin{equation}
\label{eq::lum_def}
\left(\frac{L_{1.4}}{\rm W\,Hz^{-1}}\right) = C\,\frac{4\pi}{(1+z)^{(1+\alpha)}}\,\left(\frac{D_L}{\rm Mpc}\right)^2\,\left(\frac{1.4}{3}\right)^{\alpha} \left(\frac{S_\mathrm{3}}{\rm mJy}\right)~,
\end{equation}
where $L_{1.4}$ is the 1.4 GHz K-corrected radio continuum luminosity, $C = 9.52\times10^{15}$ is the conversion factor from $\mathrm{Mpc^2\,mJy}$ to $\mathrm{W\,Hz^{-1}}$, $\alpha$ is the radio spectral index\footnote{The radio spectral index is defined as $S_{\nu} \propto \nu^{\alpha}$, where $S_{\nu}$ is the flux density at frequency $\nu$.}, $z$ is redshift, $D_L$ is the luminosity distance and $S_{\mathrm{3}}$ is the measured 3 GHz flux. We note that for 45 \% of the sources $\alpha$ was directly measured since they had a 1.4 GHz counterpart in the VLA-COSMOS joint catalogue of \cite{schinnerer10}. For the remaining sources we adopt the average spectral index, $\alpha$\,=\,-0.7, measured for galaxies in the VLA-COSMOS 3 GHz survey (Smol\v{c}i\'{c} et al., 2017a). Choosing a steeper average spectral index (e.g. $\alpha$\,=\,-0.8) would result in slightly lower IR/radio ratios and a marginally steeper redshift evolution. It would, however, still be consistent with the results presented below within 1\,$\sigma$ and affects both morphological subsamples in the same way. For a full discussion of the systematics arising from a different choice of spectral index see Sect. 4.4.1 in D17.\\
For IR sources that remain undetected in the 3 GHz map we derive an upper limit on the radio luminosity based on 5 times the local RMS noise level in the map (see Sect. 2.1 and D17 for details).

\begin{figure}
\includegraphics[width=0.48\textwidth]{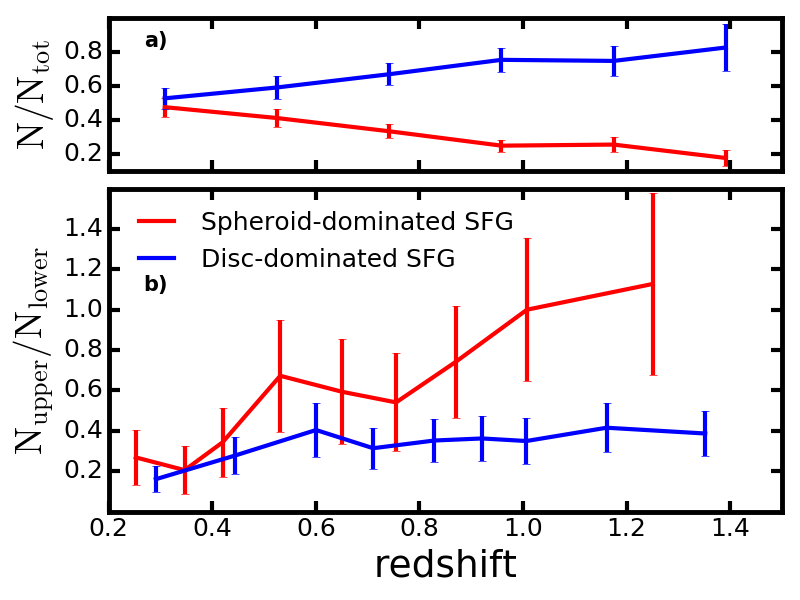}
\caption{\textbf{a)} Fractions of spheroid- (red) and disc-dominated (blue) SFGs in our star-forming sample as a function of redshift. \textbf{b)} Ratio of the number of IR-undetected ($N_{\rm upper}$) and radio-undetected ($N_{\rm lower}$) galaxies for the spheroid- and disc-dominated SFG samples in the redshift bins used in Sect. 3.2. These $N_{\rm upper}$ ($N_{\rm lower}$) objects enter the calculation of the $\overline{q}_{\mathrm{TIR}}$ values with upper (lower) limits on the IR/radio ratio, respectively. All error bars reflect 1\,$\sigma$ uncertainties from Poissonian counting statistics. Spheroid-dominated SFGs show an increasing upper limit ratio trend with redshift, whereas disc-dominated systems remain constant out to z $\sim$ 1.5.}
\label{fig::frac_vs_z}
\end{figure}

\subsection{Derivation of IR luminosities, SFRs and stellar masses with SED fitting}
\label{sect::ir_lum}

Stellar masses provided by the \textsc{magphys} code were used to physically characterise our samples (see Fig. \ref{fig::ssfr_rat}), together with star-formation rates based on the IR-luminosity of the best-fit \textsc{magphys} SED models. In this section we discuss the derivation and robustness of our IR luminosity constraints. When converting IR luminosity to SFR we apply the \cite{kennicutt98} scaling factor for a \cite{chabrier03} IMF, which is particularly suitable for star-forming galaxies and enables direct comparisons with the previous literature. SFRs calculated with this method are, on average, $\sim$0.2 dex higher than the SFR values returned by \textsc{magphys}, however, this does not impact qualitatively any of our statements based on SFRs. We further decided to adopt $L_{\rm TIR}$ derived SFR values to be consistent with the X-ray stacking analysis outlined in Sect. \ref{sect::low_lev_agn_disq}.

\subsubsection{$L_{\rm TIR}$ measurements for IR-selected galaxies}
Total IR luminosities ($L_{\rm TIR}$) were derived by integrating the best-fit SED model identified with \textsc{MAGPHYS} \citep{cunha08} between rest-frame 8 and 1000 $\mu$m. Model SEDs were fitted to photometry from the COSMOS2015 catalogue \citep[for details on the data see][]{laigle16}. In particular, in the MIR-FIR regime we used the combination of Spitzer MIPS 24 $\mu$m, plus Herschel PACS (100 and 160 $\mu$m) and SPIRE (250, 350 and 500 $\mu$m) flux measurements. For 3 (4) spheroid-dominated (disc-dominated) SFGs we were also able to use 450 and 850 $\mu$m JCMT/SCUBA-2 data \citep{casey13}.

As described in Sect. \ref{sect::joint-samp}, sources with at least one $\ge$5$\sigma$ detection in any Herschel band are included in the IR-selected sample. In the jointly-selected sample of star-forming, spheroid-dominated (disc-dominated) sources 85\% (80\%) are IR-detected (see Table \ref{tab:distr}). Within this IR-selected sample 94\% of the galaxies in both morphological sub-samples are detected at 24 $\mu$m with $S/N\,{\geq}$\,3. 67\% (69\%) of the spheroid-dominated (disc-dominated) galaxies have only one $\ge$5\,$\sigma$ Herschel measurement, 18\% (13\%) have $\ge$5$\sigma$ FIR detections in three bands, and only 1\% (1\%) have $\ge$5\,$\sigma$ fluxes across the whole FIR wavelength range. Additionally, 3--5\,$\sigma$ measurements were also included in the fitting process, where available. If, in a given band, no $\ge$3\,$\sigma$ photometry was available, we used nominal 3\,$\sigma$ PACS and SPIRE upper limits\footnote{These upper limits include both instrumental and confusion noise and are set to 5.0 (100$\mu$m), 10.2 (160$\mu$m), 8.1 (250$\mu$m), 10.7 (350$\mu$m), and 15.4 mJy (500$\mu$m).} to constrain the SED fitting, and allowed the fitting algorithm to probe fluxes below these limits via a modified $\chi^2$ calculation\footnote{Bands with upper limits contribute zero to the $\chi^2$ if the model SED falls below the upper limit. When the model SED lies above the limiting flux value(s), the excess is included as an additional term in the $\chi^2$ calculation.} following \citet{rowlands14}. While we could in principle include $\le$3\,$\sigma$ flux measurements in the SED fitting process, we chose the more conservative upper limit approach outlined above as low-$S/N$ fluxes extracted from the map are more susceptible to flux boosting due to source blending or, in the noise-dominated regime, can even take on negative values. For more details about the fitting technique and its robustness see Sect. 3 of \cite{delvecchio17}.

\subsubsection{$L_{\rm TIR}$ constraints for radio-selected galaxies}
In our analysis we treat as upper limits the IR luminosities of radio-detected sources that have no $\ge$5\,$\sigma$ measurement in any Herschel band, and are thus not part of our IR-selected sample.

The most conservative upper limits on $L_{\rm TIR}$ would result from integrating a spline through the IR flux upper limits and low-significance flux measurements. However, in addition to tracing an unphysical SED shape, this method would also ignore the available UV and optical data. We therefore decided to adopt an intermediate approach, which does not entirely neglect multi-wavelength information, but which at the same time reflects that we fundamentally wish to constrain $L_{\rm TIR}$, while only IR flux limits or poorly constrained fluxes are available in practice. Specifically, we run \textsc{MAGPHYS} on our multi-wavelength data set, with 3\,$\sigma$ upper flux limits or marginal 3--5\,$\sigma$ detections providing the flux constraints at IR wavelengths. The $L_{\rm TIR}$ values returned by \textsc{MAGPHYS} are then used as upper limits in the analysis of Sect. \ref{sect:evo_measure}. As a complementary check that upper limits on $L_{\rm TIR}$ obtained with our SED-fitting approach are neither too lax nor too strict, we compared them to $L_{\rm TIR}$ limits derived with the code SED3FIT\footnote{SED3FIT \citep{berta13} performs the SED-fitting with a modified $\chi^2$ calculation for limits. It uses a likelihood function that assigns a flux-dependent probability from the nominal upper flux limit downward, while it drops steeply above the upper limit. This statistical method ensures that model fluxes not violating the upper limits are preferred in the $\chi^2$ calculation. For details see the documentation of SED3FIT.} and find good agreement between the two methods.

Given that we employ an energy-balance SED-fitting code to constrain the IR luminosity of radio-detected sources based on both shorter wavelength data and some 3--5\,$\sigma$ FIR flux measurements, it is tempting to regard these $L_{\rm TIR}$ values as direct measurements rather than upper limits. However, since their IR SED shape is poorly constrained, this bears the risk of underestimating the true uncertainty associated with the resulting $L_{\rm TIR}$ estimates. To ensure that our findings are not driven by the ultimately somewhat subjective decision of how to statistically treat the $L_{\rm TIR}$ constraints for radio-detected sources with no $>$5\,$\sigma$ Herschel measurements, we have repeated the analysis outlined in Sect. \ref{sect:evo_measure} considering their $L_{\rm TIR}$ estimates to be measurements rather than limits. We find that this has no qualitative, and only a minimal quantitative impact on our results, and thus does not introduce any systematics that could give rise to the differences between disc- and spheroid-dominated SFGs reported below.

As a final test of the robustness of our results, we assessed the impact of poorly fitting SED models by flagging low-quality fits. Following the method presented in Appendix B of \cite{smith12}, we computed the probability of the best-fit model being consistent with our data ($p$), and identified sources with $p$\,$<$\,1\%. Excluding these sources from the analysis detailed in Sect. \ref{sect:evo_measure} did not change our results compared to our findings based on the full sample. We hence decided to retain them in order to (a) increase the overall quality of our statistics, (b) not introduce an arbitrary threshold to ``clean'' our sample and (c) follow the methodology of D17 as closely as possible, in order to enable further insight into the causes of the redshift-evolution of the IRRC found in their study.

\subsection{Measuring the infrared-radio correlation}
\label{sect:evo_measure}

The IRRC is usually characterized by the logarithmic ratio, $q_{\mathrm{TIR}}$, of the total infrared (8 \-- 1000 $\mu$m; $L_{\rm TIR}$) and 1.4 GHz radio ($L_{1.4}$) luminosities:
\begin{equation}
\label{eq::qtir}
q_{\mathrm{TIR}} \equiv \log\left( \frac{L_{TIR}}{3.75{\times}10^{12}\,\mathrm{W}} \right) - \log\left( \frac{L_{1.4}}{\mathrm{W\,Hz^{-1}}} \right).
\end{equation}
We split both the disc- and spheroid-dominated SFG samples into nine redshift bins, each bin containing an equal number of objects. We then carried out double-censored survival analysis \citep[following][]{schmitt93} to calculate the median ${q}_{\mathrm{TIR}}$ ($\overline{q}_{\mathrm{TIR}}$) values (and associated 95\% confidence intervals) for our two samples in all redshift bins.

Survival analysis reconstructs the underlying distribution assuming that all measurements (i.e. both well-defined values and upper/lower limits) are drawn from the same distribution. If, besides direct detections, only either upper or lower limits occur (i.e. the data are singly censored), the cumulative distribution function (CDF) can be constrained analytically with the Kaplan--Meier product limit estimator \citep{kaplan58}. However, if both upper and lower limits are present (i.e. the data are doubly censored), the CDF is computed by an iterative method (as described in \citealt{schmitt93} and Appendix C of \citealt{sargent10}). Our implementation of the algorithm was considered to have converged once all values of the updated CDF changed by less than 1/1000 of their value in the previous step. The estimated median $\overline{q}_{\mathrm{TIR}}$ and its error in each redshift bin was then extracted from these CDFs. We find that propagating the errors of the individual radio and IR luminosities through the analysis by resampling them a hundred times, and recalculating the CDFs, only results in a small additional uncertainty ($\sim$22\%) compared to that associated with the CDF estimating method, consistent with the assessment in D17. Thus, the errors from survival analysis dominate the error budget of $\overline{q}_{\mathrm{TIR}}$ values.

\begin{figure}
\includegraphics[width=0.46\textwidth]{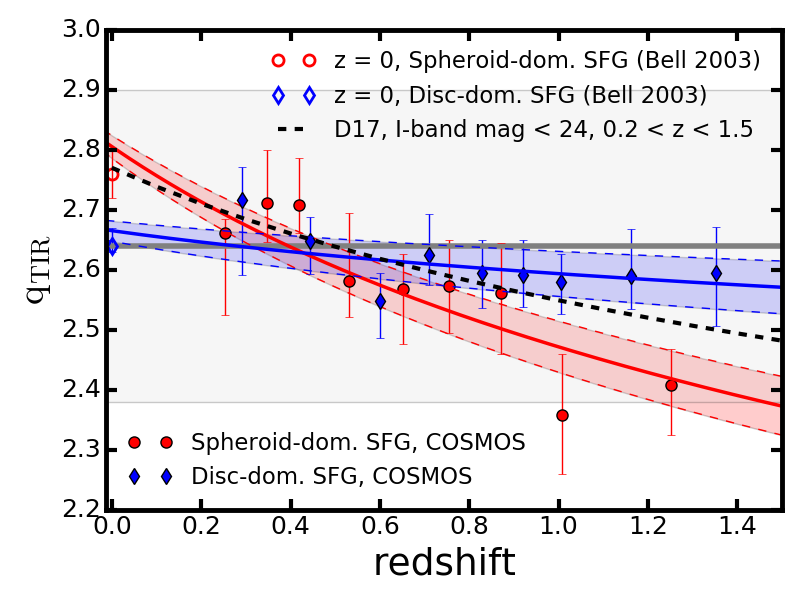}
\includegraphics[width=0.48\textwidth]{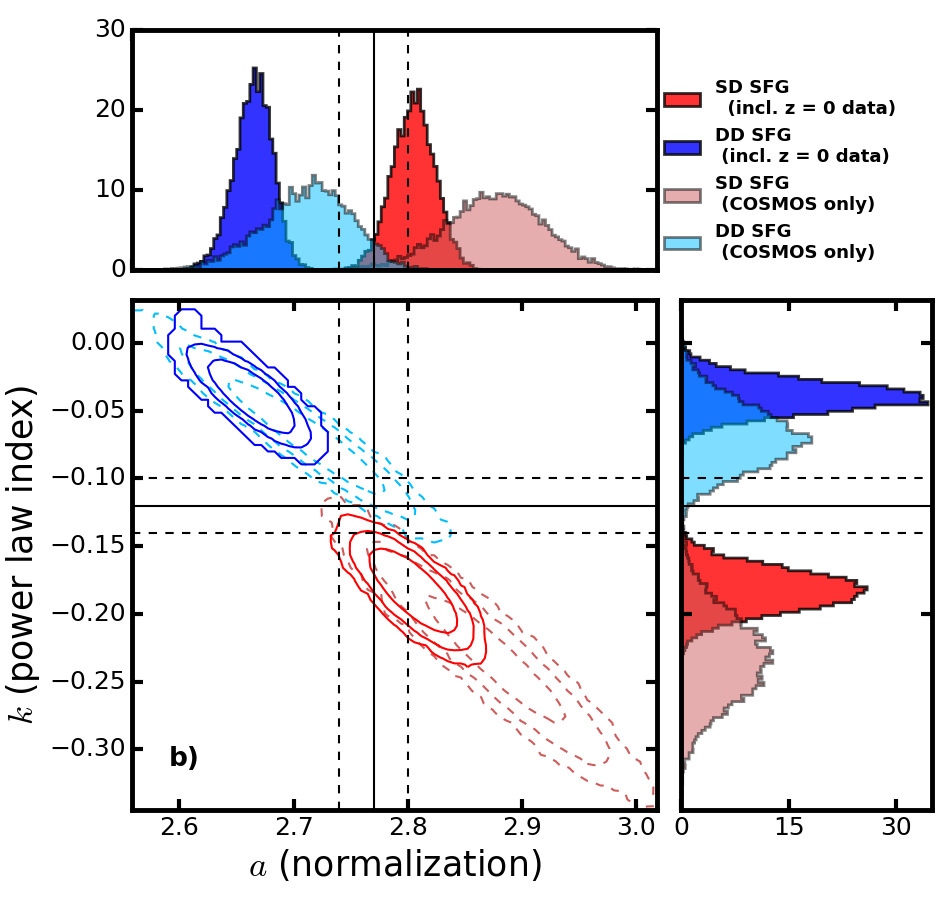}
\caption{\textbf{a)} Redshift evolution of the infrared-radio correlation, parametrized by the median IR/radio ratio $q_{\mathrm{TIR}}$, for disc- and spheroid-dominated star-forming galaxies (blue and red symbols, respectively). The $\overline{q}_{\mathrm{TIR}}$ values for COSMOS galaxies at z $>$ 0.2 (filled data points) were calculated using double-censored survival analysis. The local z = 0 measurements (open symbols) are based on a morphologically-selected sub-set of the \protect\citet{bell03} sample. Shaded regions bordered by dashed lines show the upper and lower limits of the 1 $\sigma$ confidence interval of the fit. The black dashed line is the evolutionary trend found for SFGs with I-band magnitude < 24 in the jointly-selected sample of D17 in the redshift range of 0.2 $<$ z $<$ 1.5. These cuts were imposed in order to match the selection criteria of our morphologically-selected subsamples. The dark grey horizontal line represents the local median IR/radio ratio measured by \protect\citet{bell03}, the shaded grey region its $\sim$ 0.3 scatter. \textbf{b)} 2D histogram of the parameters fitted for an evolutionary trend (eq. \ref{eq::qtir_z}) for star-forming late and spheroid-dominated galaxies in our sample, shown in blue and red, respectively. Light coloured histograms and dotted contours are the results of fits that were carried without including the local $q_{\mathrm{TIR}}$ values from \protect\citet{bell03}. Contours enclose the 3, 2 and 1 $\sigma$ confidence intervals of each distribution. The solid black lines are the fitted values for the whole star-forming population in the COSMOS field from D17. Dashed black lines represent the 1 $\sigma$ confidence interval of this fit.}
\label{fig::q_evo}
\end{figure}

The $\overline{q}_{\mathrm{TIR}}$ values calculated with survival analysis as a function of redshift are shown in Fig. \ref{fig::q_evo}a. At a qualitative level, the different redshift trends of the relative fraction of upper and lower limits observed for spheroid- and disc-dominated SFGs (see Fig. \ref{fig::frac_vs_z}b) is already a clear indication of systematically different evolution of these two morphologically distinct populations. For a quantitative confirmation, we fit the following evolutionary function to the $\overline{q}_{\mathrm{TIR}}$ measurements in all redshift bins:
\begin{equation}
\label{eq::qtir_z}
\overline{q}_{\mathrm{TIR}} = a (1+z)^k,
\end{equation}
\noindent
where $a$ and $k$ are free parameters. Errors on $a$ and $k$ were estimated by resampling the $\overline{q}_{\mathrm{TIR}}$ values using their uncertainties derived by survival analysis and refitting the newly generated datasets 10,000 times, allowing a measurement of the 1\,$\sigma$ confidence intervals. To anchor the fit, we included a $z$\,=\,0 data point based on the sample of \citet{bell03} containing structural information, which we split into spheroid- and disc-dominated SFGs following the correspondence between morphological types in the Zurich Structure \& Morphology Catalog and Hubble type described previously in Sect. \ref{sect::morph_cat}. Thus, 33 galaxies in the \citealp{bell03} sample with RC3 classification S0, S0a, Sa, Sab and Sb were assigned to the category `spheroid-dominated SFGs', and 80 RC3 types Sbc, Sc, Scd, Sd and Irr are considered disc-dominated. Omitting the local $q_{\mathrm{TIR}}$ measurements does not qualitatively change our results (see Fig. \ref{fig::q_evo}b).

\begin{figure}
\includegraphics[width=0.44\textwidth]{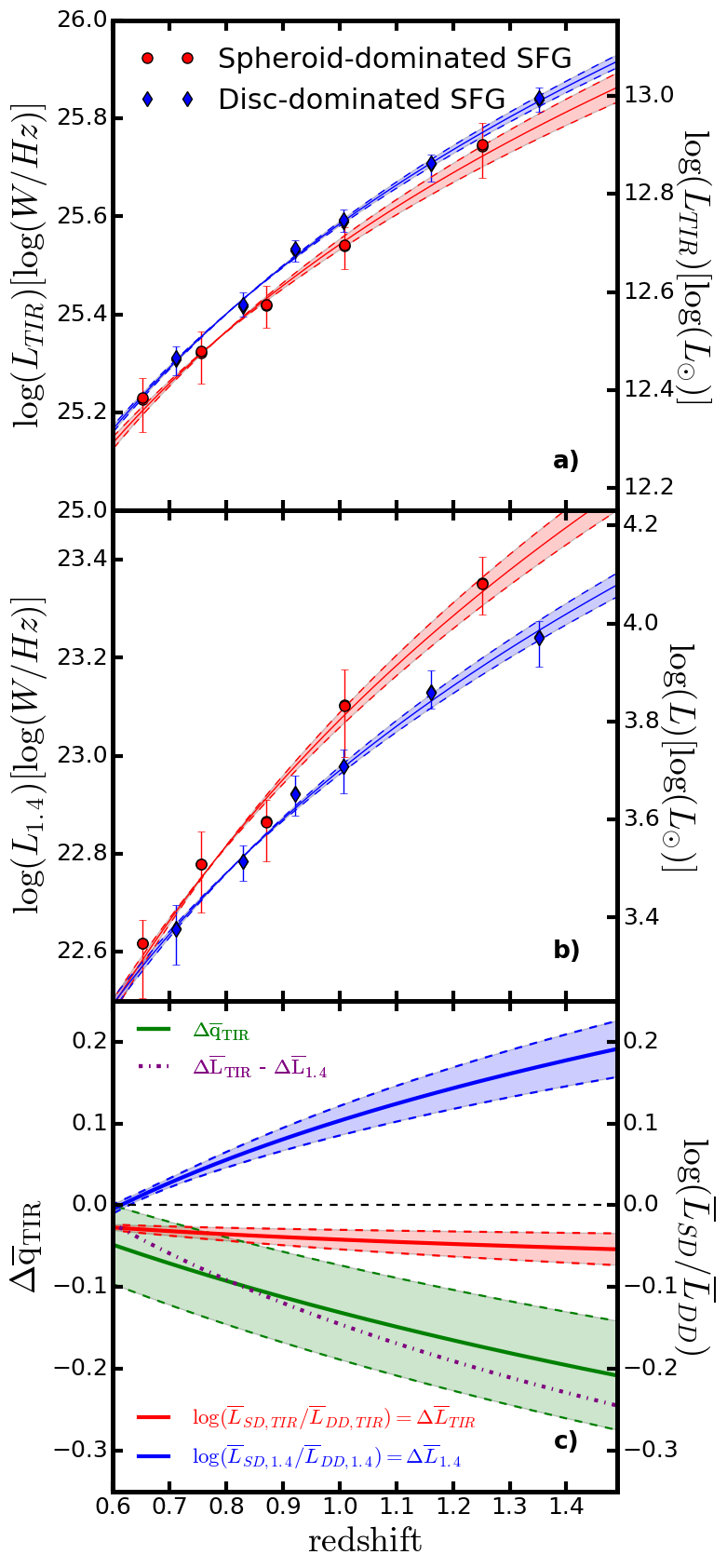}
\caption{Median total infrared (\textbf{a}) and median 1.4 GHz luminosities (\textbf{b}) in the redshift range where $\overline{q}_{\mathrm{TIR}}$ values diverge for spheroid- and disc-dominated galaxies in our sample shown in blue and red, respectively. \textbf{c)} Decomposition of the measured $\overline{q}_{\mathrm{TIR}}(z)$ difference between spheroid- and disc dominated SFGs. The green line represents the directly measured $q_{\mathrm{TIR}}(z)$ offset between spheroid- and disc-dominated SFGs from the fitted $\overline{q}_{\mathrm{TIR}}(z)$ curves in Fig. \ref{fig::q_evo} as calculated by Eq. \ref{eq::dq}. Blue and red lines are the differences in radio- and total IR-luminosities, respectively, derived from single censored survival analyses, as seen on the top two panels and Eq. \ref{eq::dL_ir_rad}. The purple line is the difference of these two trends, which also gives $\Delta \overline{q}_{\mathrm{TIR}}(z)$ (see Eq. \ref{eq::dL}.). The shaded 1 $\sigma$ confidence intervals (in appropriate colours for their corresponding lines) are calculated by propagating the errors on the previously fitted trends. It is clear that the main contributor to the different $\overline{q}_{\mathrm{TIR}}(z)$ evolution is the increasing relative radio excess of spheroid-dominated SFGs towards high redshifts.}
\label{fig::L_qdiff}
\end{figure}

We find that the median IR/radio ratios of spheroid-dominated SFGs follow a declining trend ($k\,{=}\,-0.19{\pm}0.02$), while for disc-dominated SFGs the redshift-evolution is minimal out to $z$\,=\,1.5 ($k\,{=}\,-0.04{\pm}0.01$), as shown in Fig. \ref{fig::q_evo}a. The difference between these two evolutionary trends is significant at the 7.5\,$\sigma$ level (3.3\,$\sigma$ if we exclude the $z$\,=\,0 measurements; Fig. \ref{fig::q_evo}b). In Fig. \ref{fig::q_evo}a we also show with a dashed line the best-fit evolution of $\overline{q}_{\mathrm{TIR}}$ for all star-forming galaxies (i.e. including all morphological types) that satisfy our selection criteria $i_{\rm AB}\,{\leq}\,24$ and $0.2\,{<}\,z\,{<}\,1.5$. With an evolutionary power-law index $k\,{=}\,-0.12{\pm}0.02$ it lies between the best fit relations for the spheroid and disc-dominated SFGs, as expected, and is slightly shallower than the evolutionary trend ($k\,{=}\,-0.19{\pm}0.01$) reported by D17 for the full parent sample. The main reason for difference to the measurement in D17 is that we have restricted our fit to galaxies at $0.2\,{<}\,z\,{<}\,1.5$, while D17 consider objects across all $z\,{\lesssim}\,6$.

\subsection{Differential evolution of average IR and radio brightness of spheroid- and disc-dominated galaxies}

To determine the main cause of the observed difference between spheroid- and disc-dominated star-forming galaxies, we now attempt to express the evolution of the $\overline{q}_{\mathrm{TIR}}$ value for both populations in terms of the relative evolution of their median IR and radio luminosities. To this end, we performed single-censored survival analysis on the luminosities and luminosity upper limits to constrain the according medians for the spheroid- and disc-dominated samples in all redshift bins. The redshift-evolution of the luminosities is best fit by a
\begin{equation}
\label{eq::lum_z}
\log(\overline{L}) = b z^l
\end{equation}
\noindent
model, where $\overline{L}$ is either the median rest-frame 1.4 GHz radio luminosity, or the median IR luminosity for a given redshift bin and sample. $b$ and $l$ are free parameters. The measured luminosity evolutions  and the best-fit combinations of model parameters $b$ and $l$ for spheroid- and disc-dominated star-forming galaxies are shown in Fig. \ref{fig::L_qdiff}. Radio luminosities of spheroid-dominated SFGs show an excess compared to disc-dominated SFGs above z\,$\sim$\,0.8. Their total IR luminosities have a smaller, but still significant deficit starting around the same redshift.

We define the offset between the median IR/radio ratios of spheroid- and disc-dominated SFGs as:
\begin{equation}
\Delta \overline{q}_{\mathrm{TIR}}(z) \equiv \overline{q}_{\mathrm{TIR},\mathrm{SD}}(z) - \overline{q}_{\mathrm{TIR}, \mathrm{DD}}(z),
\label{eq::dq}
\end{equation}
\noindent
where $\overline{q}_{\mathrm{TIR},\mathrm{SD}}(z)$ and $\overline{q}_{\mathrm{TIR}, \mathrm{DD}}(z)$ are the $\overline{q}_{\mathrm{TIR}}(z)$ sample median trends fitted between 0\,<\,$z$\,<\,1.5 for spheroid- and disc-dominated SFGs, respectively. We also define:
\begin{equation}
\Delta \overline{L}_{\{1.4,\mathrm{TIR}\}}(z) \equiv \log \overline{L}_{\{1.4,\mathrm{TIR}\},\,SD}(z) - \log \overline{L}_{\{1.4,\mathrm{TIR}\},\,DD}(z),
\label{eq::dL_ir_rad}
\end{equation}
\noindent
where $\overline{L}_{\{1.4,\mathrm{TIR}\},\,SD}$ is either the median 1.4 GHz or IR luminosity of spheroid-dominated SFGs and $\overline{L}_{\{1.4,\mathrm{TIR}\},\,DD}$ is the corresponding median luminosity for disc-dominated SFGs. Then from Eq. \ref{eq::qtir} and Eq. \ref{eq::dL_ir_rad}  we see that:
\begin{equation}
\Delta \overline{q}_{\mathrm{TIR}}(z) = \Delta \overline{L}_{\mathrm{TIR}}(z) - \Delta \overline{L}_{1.4}(z).
\label{eq::dL}
\end{equation}

This  decomposition of the measured $\overline{q}_{\mathrm{TIR}}(z)$ difference between spheroid- and disc dominated SFGs is shown in Fig. \ref{fig::L_qdiff}. It suggests that the main cause of the different observed $q_{\mathrm{TIR}}$ trends is an increasing radio excess in spheroid-dominated SFGs compared to disc-dominated SFGs above $z$\,$\sim$\,0.8, with contribution from a smaller IR deficit of spheroid-dominated SFGs in the same redshift range. This hints at additional AGN-related radio emission, possibly from small-scale jet-activity at higher redshifts in bulge-dominated systems. We will explore this point further in Sect. \ref{sect::dq_discussion}.

\section{Discussion}

\subsection{Minimal evolution of the IR-radio correlation for disc-dominated star-forming galaxies}

As shown in Fig. \ref{fig::frac_vs_z}b, spheroid- and disc-dominated SFGs show different censoring patterns (i.e. different balance of direct measurements and upper/lower limits) for their IR/radio ratios above $z$\,$\sim$\,0.8. This in itself already points to a differential evolution of sample medians $\overline{q}_{\mathrm{TIR}}$ towards higher redshifts. As found by doubly-censored survival analysis, the radio-infrared ratios of the disc-dominated SFG sample show virtually no evolution out to $z$\,$=$\,1.5 and are in almost all redshift bins consistent with the locally measured median $q_{\mathrm{TIR}}$ value of \citet{bell03} (Fig. \ref{fig::q_evo}). The fact that the median IR/radio ratio is nearly constant for that class of galaxies which effectively represents the `proto-typical' (i.e. disc-like) star-forming object, suggests that -- when applied to purely star-forming systems/regions -- radio synchrotron emission traces star formation in the same way over the entire redshift range 0\,<\,$z$\,<\,1.5. We note that, above $z$\,=\,1, our morphological selection may be somewhat biased towards classification into disc-dominated SFGs as resolution effects and decreasing pixel signal-to-noise in HST images can reduce the measured concentration index (and hence bulge-to-disc ratio) of the more distant galaxies. As a consequence, we expect that, if anything, the disc-dominated SFG sample presumably contains some galaxies which would be classified as spheroid-dominated SFGs in noise-free HST images, rather than the other way around. If there is indeed a systematic difference between the IR/radio ratios $q_{\mathrm{TIR}}$ of spheroid and disc-dominated systems as suggested by our analysis in Sect. 3.2, correcting for this morphological classification bias would further flatten the evolutionary trend for the disc-dominated population. We note that removing the 3\,$\sigma$ radio excess outliers from the disc-dominated sample also flattens their $\overline{q}_{\mathrm{TIR}}(z)$ trend further ($k$\,=\,-0.02$\pm$0.02). The flagging and removal of outlier ${q}_{\mathrm{TIR}}$ values are described in Sect. \ref{sect::dq_discussion}. Starburst objects (i.e. $\log(\mathrm{sSFR/sSFR_{MS}}) > 0.6$) are more abundant among disc-dominated systems (as seen in Fig. \ref{fig::ssfr_rat}a). However, removing them from both samples does not change our results. In fact, we note that there is no significant change in $\overline{q}_{\mathrm{TIR}}$ as a function of MS offset across the entire redshift range \citep[in agreement with the findings of][]{magnelli15}.

\citet{lacki10b} defined a `critical redshift' ($z_{\mathrm{crit}}$) for their models, by which the radio luminosity is suppressed by a factor of 3 compared to $z$\,=\,0 due to increasing inverse Compton losses off the cosmic microwave background (CMB). This would lead to an increase in $q_{\mathrm{TIR}}$ of $\sim$0.5\,dex. In their model $z_{\mathrm{crit}}$ is ultimately determined by the SFR surface density ($\Sigma_{\mathrm{SFR}}$) of a galaxy. In order to compare our findings to theoretical predictions, we derived the median $\Sigma_{\mathrm{SFR}}$ of our SFG sample using optical half-light radii from the Zurich Structure \& Morphology Catalog and SFRs calculated from the total IR luminosities (see Sect \ref{sect::ir_lum} for details). We note that it is improbable that this approach should strongly overestimate $\Sigma_{\mathrm{SFR}}$ as, e.g., \citet{nelson16} and \citet{rujopakarn16} find approximately similar sizes for SF activity and stellar mass in distant SFGs, and especially since HST/ACS F814W filter samples blue rest-frame emission at $z\,{>}\,1$. Given that the bulk of our galaxies in both samples lie on the SF MS (see Fig. \ref{fig::ssfr_rat}a), we convert $\Sigma_{\mathrm{SFR}}$ into $z_{\mathrm{crit}}$ using the formula for normal galaxies from the simplest model\footnote{Their formulae for different starburst models, SFR laws and feedback mechanisms predict higher $z_{\mathrm{crit}}$ values for the same $\Sigma_{\mathrm{SFR}}$, hence the $z_{\mathrm{crit}}$ values given in text should be considered as lower limits. However, a higher $z_{\mathrm{crit}}$ would qualitatively not change our conclusions.} in \citet{lacki10b}. The median $\Sigma_{\mathrm{SFR}}$ of $\sim$0.08\,$\mathrm{M_{\odot}\,kpc^{-2}\,yr^{-1}}$ derived for our disc-dominated SFGs yields $z_{\mathrm{crit}}\,{\sim}$\,2.2, i.e. \citet{lacki10b} do not predict a significant change in $\overline{q}_{\mathrm{TIR}}$ for objects of this $\Sigma_{\mathrm{SFR}}$ across our redshift range, in agreement with our data. However, their model finds a very similar $z_{\mathrm{crit}}$ of 2.3 for spheroid-dominated SFGs due to their similar median $\Sigma_{\mathrm{SFR}}$ of $\sim$0.09\,$\mathrm{M_{\odot}\,kpc^{-2}\, yr^{-1}}$, which in this case is in discrepancy with the decreasing $\overline{q}_{\mathrm{TIR}}(z)$ we find. It is interesting to note that, should the sizes of star-forming regions be significantly smaller than assumed here, as suggested by the 10\,GHz imaging results in \citet{murphy17}, the correspondingly higher $\Sigma_{\mathrm{SFR}}$ would push $z_{\mathrm{crit}}$ to even higher redshifts.

\subsection{Differential redshift-evolution of $\overline{q}_{\mathrm{TIR}}$ depending on disc galaxy type -- possible explanations}
\label{sect::dq_discussion}

If we omit disc galaxies with intermediate bulge-to-disc ratios (i.e. ZEST type 2.1 and 2.2) from the spheroid- and disc-dominated SFG samples, their $\overline{q}_{\mathrm{TIR}}(z)$ power-law indices are in even stronger contrast (k$=$-0.22$\pm$0.06 and k$=$0.03$\pm$0.02, respectively). In the following section we will investigate possible causes of this morphology-dependent change in their evolutionary trends.

\subsubsection{Low-level AGN contamination}
\label{sect::low_lev_agn_disq}

The lower $\overline{q}_{\mathrm{TIR}}$ values of spheroid-dominated SFGs at redshifts $z\,{\gtrsim}$\,0.8 are mainly due to their relative radio excess compared to disc-dominated SFGs at high $z$ (Fig. \ref{fig::L_qdiff}). This could be the result of a radio component related to weak AGN activity in bulgy SFGs. We note that, as shown in Sect. \ref{sect::agn_id}, a four times higher fraction of systems were flagged and removed as AGN hosts in the spheroid-dominated sample, compared to the disc-dominated sample. If the measured radio excess of spheroid-dominated disc galaxies were entirely due to residual AGN activity (i.e. if the SF-related radio and infrared emission from these objects followed the local IRRC as is observed for the disc-dominated systems), then we would infer that at $z\,{\sim}\,1.5$ on average $\sim$\,60\% of their total observed radio emission is contributed by AGN-related processes (see Fig. \ref{fig::L_qdiff}.), which escaped detection by the full set of standard AGN identification methods described in Sect. \ref{sect::agn_id}.

Alternatively, the additional average radio emission could be the consequence of a relatively small number of strong radio excess sources that appear in our data as outliers in the $q_{\mathrm{TIR}}$ distribution in each redshift bin. The CDFs in each bin derived by survival analysis allow us to select and remove such sources from each of our sub-samples. Assuming that the $q_{\mathrm{TIR}}$ distribution is well described by a Gaussian distribution, we computed the $\sigma$ of the distribution, and removed 3\,$\sigma$ outliers below the median in all our redshift bins. We then repeated our survival analysis on these ``cleaned'' samples. The resulting $\overline{q}_{\mathrm{TIR}}$ trends are consistent with the ones measured on the full sample within 1\,$\sigma$. In fact, applying a more stringent 2\,$\sigma$ lower $q_{\mathrm{TIR}}$ cut, where $\sigma$ was derived only using $q_{\mathrm{TIR}}$ values higher than the median, yielded the same slopes for the trends. This implies that the measured evolutions of the $\overline{q}_{\mathrm{TIR}}$ values are not driven by low $q_{\mathrm{TIR}}$ outliers for both the spheroid- and disc-dominated SFG samples.

To further investigate potential residual AGN contamination of our SFG samples we searched for small scale radio emission associated with our galaxies by cross-matching the spheroid- and disc-dominated SFG samples at $z\,{>}\,0.8$, where $\overline{q}_{\mathrm{TIR}}$ trends start to diverge, with the 1.4 GHz VLBA-COSMOS catalogue (Herrera-Ruiz, submitted., l0 $\mu$Jy sensitivity in the central part of the field). We found 9 (1.6\%) and 15 (0.9\%) counterparts for spheroid- and disc-dominated SFGs, respectively. This rules out any significant contamination from sources with $L_{1.4}\,{>}\,1 - 6{\times}10^{23}$\,W\,Hz$^{-1}$ radio cores (at  $z\,{=}\,0.8$ and 1.5, respectively) in both samples at this redshift range. However, if the 0.1 -- 0.2 dex excess radio luminosity is due to small-scale AGN-related emission in spheroid-dominated galaxies at z $>$ 1 (as seen on Fig. \ref{fig::L_qdiff}.), then such components fall below the detection limit of the VLBA sample by a factor of $\sim$\,3. Thus the low matching fraction does not rule out the possibility of a non-SF related radio component in the bulge of spheroid-dominated galaxies.\\
Unobscured AGN activity could also manifest itself as centrally concentrated emission in our high-resolution, rest-frame optical HST images. If this were preferentially the case for SFGs classified as spheroid-dominated, we might expect a stronger evolution of average concentration indices (CIs) for these systems at $z\,{\gtrsim}\,0.8$, compared to disc-dominated SFGs. Using the Zurich Structure \& Morphology Catalog we calculated the median CIs of the spheroid- and disc-dominated SFG samples, but find that the median CI of spheroid-dominated SFGs is consistently $\sim$20\% larger than that of disc-dominated SFGs across the whole redshift range. This lack of differential evolution between the two samples provides further evidence against an increasing number of highly-concentrated spheroid-dominated systems, that could be linked to higher AGN contamination.

\begin{figure}
\includegraphics[width=0.5\textwidth]{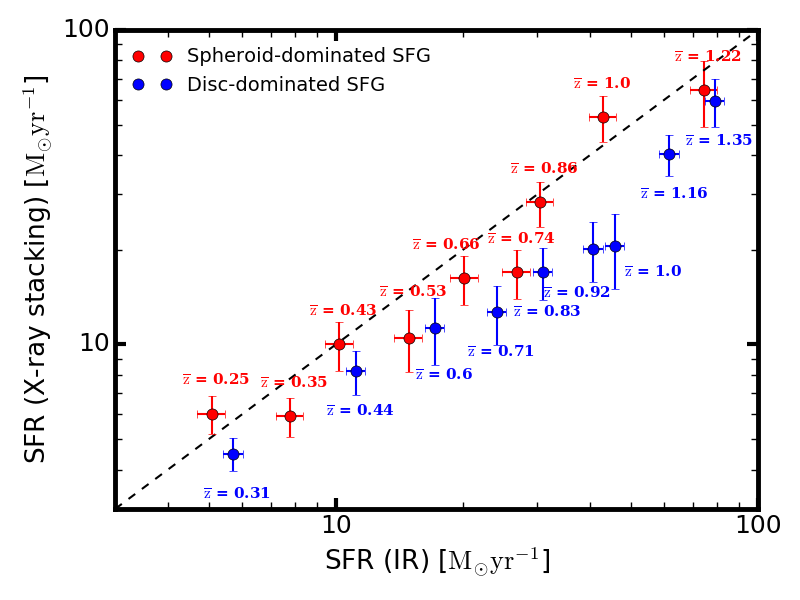}
\caption{A comparison of star formation rates derived from X-ray stacking and Herschel based IR flux averaging for both SFG samples. Dashed line represents the 1:1 relation.}
\label{fig::x_ray_stack}
\end{figure}

As mentioned in Sect. \ref{sect::agn_id}., none of our SFGs have X-ray luminosities ($\mathrm{L_X}$) > $10^{42}$ erg s$^{-1}$. As a final test to gauge AGN contamination at high-redshifts, we carried out X-ray flux stacking in both the spheroid- and disc-dominated SFG samples for comparison. We used \textsc{cstack}\footnote{Written by Takamitsu Miyaji, and available at \url{http://lambic.astrosen.unam.mx/cstack}}, a publicly-available tool for stacking, to stack soft ([0.5 - 2] keV) and hard band ([2 - 8] keV) Chandra images of our sources in each redshift bin. The resulting stacked count rates were converted to X-ray luminosities with a 1.4 slope power law spectrum, as found for the X-ray background \citep[e.g.][]{gilli07}. We then converted the X-ray luminosities to SFRs with the calibration given by \citet{symeonidis14}. In the same redshift bins we calculated average SFRs from IR luminosities, using the simple conversion of \citet{kennicutt98} with a \citet{chabrier03} IMF in order to facilitate the comparison with other SFR estimates adopted in the literature (see Sect. \ref{sect::ir_lum}). These are less sensitive to AGN related emission, thus provide a reference for finding X-ray excess in our samples. Fig. \ref{fig::x_ray_stack}. shows the comparison between X-ray and IR-derived SFRs for both spheroid- and disc-dominated SFGs. We found a small systematic X-ray deficit in the disc-dominated SFG sample, most likely due to the fact that they are not representative of the galaxy population used for the calibration of the $\mathrm{SFR_{X}}$ - $\mathrm{SFR_{IR}}$ relation in \citet{symeonidis14}. If we were to rescale it to better fit our disc-dominated SFGs, the spheroid-dominated sample above $z\,{=}\,0.8$ would end up having a $\sim$0.2 dex X-ray excess. Considering that the $\mathrm{SFR_{X}}$ - $\mathrm{SFR_{IR}}$ calibration also has a $\sim$0.2 dex scatter, it would still not suggest a strong X-ray excess in this redshift range.

In summary, we have found no conclusive signs of a higher AGN contamination in our spheroid-dominated SFG sample above redshift $\sim$ \,1, however, with available data we cannot exclude this scenario. Should the excess radio emission of spheroid-dominated SFGs be at least partly AGN-related, it could be viewed as the indirect signature of the build-up of black hole-bulge mass correlation at small growth rates \citep[see also][]{mullaney12}.

\subsubsection{Other contributing factors}

If the radio excess observed for $z\,{>}\,0.8$ spheroid-dominated SFGs is not due to a significant AGN contamination, another possibility is a flattening radio spectral slope trend with redshift. A constant $\alpha\,{=}\,-0.7$, would overestimate the rest-frame 1.4 GHz luminosity derived from 3 GHz fluxes (see Eq. \ref{eq::lum_def}). Using the radio excess curve for spheroid-dominated systems in Fig. \ref{fig::L_qdiff} we calculate that the median radio spectral index of spheroid-dominated SFGs would have to increase from $\alpha\,{=}\,-0.7$ at $z\,{=}\,0.8$ to $\alpha\,{=}\,-0.45$ at $z\,{=}\,1.5$ in order to cancel the radio excess and produce $\overline{q}_{\mathrm{TIR}}$ values consistent with the locally measured one at all redshifts. This is inconsistent both with the trend for decreasing 1--3\,GHz spectral index values reported by D17, as well as with the median radio spectral slopes of disc- and spheroid-dominated systems we actually measure, and which are consistent with each other across the entire redshift range. On the other hand, \citet{murphy17} find evidence for a flatter ($\alpha\,{>}\,-0.6$) average radio spectral index over the wider frequency baseline 1--10\,GHz. Given the large spectral index dispersion of 0.35\,dex found by \citet{murphy17} it is thus possible that, at least for some galaxies, spectral indices may approach values required for reducing the measured radio excess. However, we also note that, while restricted to low-redshift samples (e.g. \citealp{niklas97} and \citealp{marvil15}), previous studies do not find evidence for differently shaped radio spectra for star-forming galaxies with different optical morphologies.

While the overall differential evolution implied by the fitted best fit relations (Eq. \ref{eq::qtir_z}.) is significant at the 7.5\,$\sigma$ level, the actual data underpinning this trend does not display as smooth an evolution as the best-fit power law. This could be interpreted as a signature of multiple different factors being at play. At low redshifts ($z\,{<}\,0.3$), using local $\overline{q}_{\mathrm{TIR}}$ estimates from \citet{bell03}, we find that spheroid-dominated systems have higher $\overline{q}_{\mathrm{TIR}}$ values than disc-dominated galaxies, in qualitative agreement with the findings of, e.g., \citet{nyland17}. This could arise from an IR-excess related to enhanced cirrus emission in spheroid-dominated systems linked to old stellar populations rather than SF activity. As a test, we calculated median cold and warm dust component temperatures fitted with \textsc{magphys} in each redshift bin. During the SED fitting \textsc{magphys} decomposes the FIR part of the SED into a warm component, related to birth clouds where SF activity occurs, and a cold component, representing the ISM heated by an on average older stellar population. We found that the cold component of spheroid-dominated systems is marginally warmer (by 1 K or $\sim$\,4\%) than in disc-dominated SFGs below $z\,{=}\,0.7$. However, calculating median infrared/radio ratios using the cold or the warm component yields redshift evolution slopes consistent within 1\,$\sigma$ with $\overline{q}_{\mathrm{TIR}}$ slopes, further suggesting that the difference is mainly driven by radio emission rather than a changing balance between the cold and warm (i.e. star-formation related) dust emission in spheroid- and disc-dominated SFGs. We caution that, due to low detection rates 350 and 500\,${\mu}$m, detecting a cold \citep[dust temperature 20 K as found by e.g.][]{bianchi17} cirrus component is challenging with our data, in particular at $z\,{>}\,1$. While it is thus possible that such a component could be missed (and IR-luminosities thus underestimated), widespread occurrence of cirrus emission strong enough to flatten the $\overline{q}_{\mathrm{TIR}}(z)$ trend for spheroid-dominated SFGs would presumably result in larger SPIRE detection fractions for this population. This is, however, not what we observe, such that it is unlikely that missed cirrus emission is responsible for the observed difference between the evolutionary trends measured for spheroid- and disc-dominated SFGs in Sect. \ref{sect:evo_measure}.

In conclusion, one could envisage a scenario where a combination of excess IR due to cirrus emission at low redshifts and an AGN contribution at high redshifts (with a transition around $z\,{\sim}\,0.8$) jointly drive the $\overline{q}_{\mathrm{TIR}}$ trend for the spheroid-dominated SFGs. Regardless of the physical processes regulating the $\overline{q}_{\mathrm{TIR}}$ evolution, from an empirical point of view, the observed differential trends found in this study highlight the importance of ancillary data, such as morphological information, for the process of converting observed radio fluxes into SFR measurements, especially for galaxies at redshifts above 1. While radio emission apparently traces SFR much in the same way at $z\,{\sim}\,1$ as it does at $z\,{\sim}\,0$ (as suggested by the near constancy of the median IR/radio ratios of massive, disc-dominated SFGs across this redshift range), it may nevertheless be more appropriate to adopt a recipe involving a redshift-dependent IR/radio ratio $\overline{q}_{\rm TIR}(z)$ (see, e.g., eq. 4 in D17)
\begin{equation}
SFR \propto10^{\overline{q}_{\rm TIR}(z)}\,L_{1.4}
\end{equation}
when dealing with a purely SF-selected sample, since this kind of prescription can statistically account for the average fraction of radio emission unrelated to ongoing SF activity. However, we find that this correction is preferentially needed in spheroid-dominated systems, implying that combining morphological indicators with radio data can increase the accuracy of radio-based SFR estimates when this additional information is available.

\section{Summary}

With the combination of infrared data from Herschel Space Observatory and new, high sensitivity Karl G. Jansky Very Large Arrany (VLA) 3 GHz observations in the COSMOS field, and morphological classification from the Zurich Structure \& Morphology Catalog, we studied the redshift dependence of the infrared/radio ratio of spheroid- and disc-dominated SFGs on the star-forming main sequence out to a redshift of 1.5. We found that the median IR/radio flux ratio $\overline{q}_{\mathrm{TIR}}$ of disc-dominated galaxies shows virtually no evolution, in agreement with e.g. the model of \citet{lacki10b}. This suggests, that calibrations of radio luminosity as a SFR tracer based on local galaxies remain valid out to $z\,{\sim}\,1.5$ for `pure' star-forming systems. It also implies that disc-dominated galaxies may be the most suitable laboratories for studying the evolution and physics of the infrared-radio correlation, as in these systems radio and IR emission are linked to star formation in the most straightforward way. Spheroid-dominated SFGs, on the other hand display a decreasing trend with a slope of $-0.19{\pm}0.02$, consistent with most recent literature on the evolution of the IRRC for star-forming galaxies in general. A comparison of total infrared and radio luminosities between these two morphologically distinct sub-samples of SFGs revealed that the low $q_{\mathrm{TIR}}$ values above $z\,{\sim}\,0.8$ for spheroid-dominated SFGs are mainly the result of their $\sim$10--60\% radio luminosity excess relative to disc-dominated systems. This could hint at AGN activity at radio frequencies that did not reveal itself clearly in standard AGN diagnostics, which were initially used to identify and remove AGNs from both samples.

The fact that morphologically distinct samples of galaxies follow different redshift trends implies that future high-resolution and high-sensitivity surveys aiming to use radio continuum observations as a star formation tracer will strongly benefit from ancillary morphological data for the final analysis step of converting radio luminosities into accurate SFR estimates. Using a more nuanced calibration of this kind will be possible for high-redshift studies building on the synergies between radio surveys with, e.g., the Square Kilometre Array and high-resolution optical information obtained with Euclid \citep{ciligeli15}. From a theoretical perspective it highlights the complex interplay of physical processes contributing to galaxy-integrated measurements and the need to disentangle these for a more thorough understanding of the IRRC.

\section*{Acknowledgements}

We thank the anonymous reviewer for a helpful report that allowed us to improve the manuscript. We thank Anna Cibinel, Steven Duivenvoorden, Philipp Lang, Maurilio Pannella, Isabella Prandoni, Veronica Strazzullo, David Sullivan and Sune Toft for useful discussions. DM acknowledges support from the Science and Technology Facilities Council (grant number ST/M503836/1). MTS was supported by a Royal Society Leverhulme Trust Senior Research Fellowship (LT150041). JD, ID, VS, MN acknowledge support from the European Union's Seventh Frame-work program under grant agreement 337595 (ERC Starting Grant, ``CoSMass"). NHR acknowledges support from the Deutsche Forschungsgemeinschaft through project MI 1230/4-1. This publication has received funding from the European Union's Horizon 2020 research and innovation programme under grant agreement No 730562 [RadioNet]. This research made use of \textsc{astropy}, a community-developed core \textsc{python} package for Astronomy \citep{astropy13} Much of the analysis presented here was carried out in the Perl Data Language \citep[PDL;][]{glazebrook97} which can be obtained from \url{http://pdl.perl.org}.

%%%%%%%%%%%%%%%%%%%%%%%%%%%%%%%%%%%%%%%%%%%%%%%%%%

%%%%%%%%%%%%%%%%%%%% REFERENCES %%%%%%%%%%%%%%%%%%

% The best way to enter references is to use BibTeX:

\bibliographystyle{mnras}
\bibliography{irrc_zagreb}

%%%%%%%%%%%%%%%%%%%%%%%%%%%%%%%%%%%%%%%%%%%%%%%%%%

%%%%%%%%%%%%%%%%% APPENDICES %%%%%%%%%%%%%%%%%%%%%

%\appendix

%\section{Some extra material}

%If you want to present additional material which would interrupt the flow of the main paper,
%it can be placed in an Appendix which appears after the list of references.

%%%%%%%%%%%%%%%%%%%%%%%%%%%%%%%%%%%%%%%%%%%%%%%%%%

% Don't change these lines
\bsp	% typesetting comment
\label{lastpage}
\end{document}